\documentclass[sigconf,screen]{acmart}

\usepackage{booktabs}
\usepackage{multirow}

\AtBeginDocument{\providecommand\BibTeX{{\normalfont B\kern-0.5em{\scshape i\kern-0.25em b}\kern-0.8em\TeX}}}

\copyrightyear{2026}
\acmYear{2026}
\setcopyright{cc}
\setcctype{by}
\acmConference[CHI '26]{Proceedings of the 2026 CHI Conference on Human Factors in Computing Systems}{April 13--17, 2026}{Barcelona, Spain}
\acmBooktitle{Proceedings of the 2026 CHI Conference on Human Factors in Computing Systems (CHI '26), April 13--17, 2026, Barcelona, Spain}
\acmPrice{}
\acmDOI{10.1145/3772318.3790924}
\acmISBN{979-8-4007-2278-3/2026/04}

\begin{document}

\title{Risk, Data, Alignment: Making Credit Scoring Work in Kenya}

\author{Daniel Mwesigwa}
\email{dm663@cornell.edu}
\orcid{0000-0002-2735-4875}
\affiliation{\institution{Cornell University}
  \streetaddress{Gates Hall, 107 Hoy Rd}
  \city{Ithaca}
  \state{NY}
  \country{United States}}
  \postcode{14853}

\author{Steven J. Jackson}
\email{sjj54@cornell.edu}
\orcid{0000-0002-4426-1320}
\affiliation{\institution{Cornell University}
  \streetaddress{Gates Hall, 107 Hoy Rd}
  \city{Ithaca}
  \state{NY}
  \country{United States}}
  \postcode{14853}

\author{Christopher Csikszentmihalyi}
\email{cpc83@cornell.edu}
\orcid{0000-0001-8288-5268}
\affiliation{\institution{Cornell University}
  \streetaddress{Gates Hall, 107 Hoy Rd}
  \city{Ithaca}
  \state{NY}
  \country{United States}}
  \postcode{14853}

\begin{abstract}

Credit scoring is an increasingly central and contested domain of data and AI governance, frequently framed as a neutral and objective method of assessing risk across diverse economic and political contexts. Based on a nine-month ethnography of credit scoring practices in Nairobi, Kenya, we examined the sociotechnical and institutional work of data science in digital lending. While established regional telcos and banks are leveraging proprietary data to develop digital loan products, algorithmic credit scoring is being transformed by new actors, techniques, and shifting regulations. Our findings show how practitioners construct alternative data using technical and legal workarounds, formulate risk through multiple interpretations, and negotiate model performance via technical and political means. We argue that algorithmic credit scoring is accomplished through the ongoing work of alignment that stabilizes risk under conditions of persistent uncertainty, taking epistemic, modeling, and contextual forms. Extending work on alignment in HCI, we show how it operates as a \textit{two-way} translation, where models are made “safe for worlds” while those worlds are reshaped to be “safe for models.”

 \end{abstract}

\ccsdesc[500]{Human-centered computing~Empirical studies in HCI}
\ccsdesc[300]{Human-centered computing~Theory, models and concepts}

\keywords{Algorithmic credit scoring, digital lending, alternative data, ethnography, data science, machine learning practice, Kenya}


\maketitle

\section{Introduction}

In Nairobi, Kenya, new models of algorithmic credit scoring, often developed using techniques of artificial intelligence (AI), machine learning (ML), and data science, are emerging in response to growing consumer demand for credit and efforts by financial technology companies to expand their market reach \cite{munyendo_desperate_2022,njuguna_poster_2021,park_privacy_2023,qureshi_discreating_2021,shema_effective_2019,speakman_three_2018}. Kenya leads Africa in digital lending: the World Bank’s 2025 Global Findex report \cite{klapper_global_2025} shows that 32\% of Kenyan adults borrow from mobile-money providers (25\% borrow exclusively this way); and 86\% of borrowers use mobile money to meet needs such as school fees and everyday consumption. Despite the significant uptake of digital financial services, fintechs in Kenya have been accused of high interest rates (above 100\% if annualized), aggressive debt collection tactics such as public shaming, and consumer privacy violations \cite{odwyer_algorithms_2018,kiruga_this_2020,munyendo_desperate_2022}. For their part, borrowers have adopted a range of practices to navigate their credit obligations, including changing SIM (identity) cards or refinancing digital loans by borrowing elsewhere \cite{park_privacy_2023,thieme_hustle_2025,waithira_soulless_2024}. The result has been a growing default rate on digital loans, which has soared as high as 40\% \cite{muiruri_heat_2025}. Digital credit providers, including telcos, banks, and startups, are now tackling these defaults and “risk frontiers” by doubling down on techniques such as algorithmic credit scoring and factoring in new sources of data about local credit cultures and patterns of consumer behavior. These technology-centered responses, while seemingly new, have been envisioned as part of the financial inclusion toolkit for over two decades. For example, data sources such as mobile money and airtime transaction logs have been explored for their potential to enhance algorithmic credit scoring, particularly where traditional credit reference bureau coverage is partial or limited \cite{world_bank_doing_2009}. What has changed is the scale of the use of novel ML techniques and datasets, and the number of consumers affected by them.

Credit scoring is a prime example of algorithmic risk management, highlighting how algorithms are developed and embedded in high-stakes decision domains \cite{kiviat_credit_2019,lauer_creditworthy_2017}. Indeed, the use of algorithmic techniques and ML technologies for predictive purposes has applications beyond credit risk assessments, including criminal justice \cite{angwin_machine_2016}, social welfare \cite{saxena_human-centered_2020}, and healthcare \cite{chancellor_who_2019}. A growing number of studies in HCI (human-computer interaction) and related fields have provided valuable insights into automated decision systems and their social impacts. For example, \citet{barocas_big_2016} have shown how big data’s far-reaching capabilities index proxy features of protected attributes such as gender, race, and religion,  reproducing historical biases and social inequities through automated decision systems. Because companies closely guard their algorithms and tools as proprietary, it has been challenging to investigate the full range of assumptions, design choices, and values underlying these technologies \cite{hurley_credit_2016}; although AI audits and community-centered approaches to research have surfaced real user concerns and forged concrete paths for accountability \cite{cunningham_cost_2022,kotut_terms_2025,lee_balance_2025,smith_many_2023}. Other work, building on established traditions of workplace ethnography in laboratory settings, has explored the core organizational and collaborative practices that shape credit scoring, and how these practices encounter and are constituted in different social worlds. Such studies have generated in-situ accounts of how values such as trust \cite{passi_trust_2018} and transparency \cite{fahimi_friction_2023} are defined and operationalized in the everyday work of data science, and brought into alignment within organizational contexts, shifting policy mandates, and wider social interests and values.

Beyond these typically Western-centered sites and contexts, organizational studies of data science practice in much of the majority world have been limited in scope and dominated by an access or ``inclusionist'' paradigm oriented to the extension of digital credit (and other financial services) under conditions that are understood to be data (and credit) poor \cite{amrute_primer_2022,arora_bottom_2016,natale_global_2025}. For example, Shema’s \cite{shema_effective_2019} study of the efficacy of privacy-preserving credit scoring in a leading telco in Rwanda reported that comparable performance metrics can be attained by using limited data (i.e., airtime recharge) rather than expansive, privacy-infringing consumer data. In a different context, \citet{speakman_three_2018} deployed a phone-based credit product in an East African country that lacked relevant training data by leveraging datasets from another country in the region with a fully developed digital lending ecosystem. These interventions have generated useful insights about what works technically, but not how such systems are built in practice, particularly within the broader and distinctive dynamics of credit scoring in Kenya: the unevenness of technology competition, including Safaricom’s near monopoly (95+\%) of telecommunication and financial technology data \cite{mazer_fast_2024}; unique credit cultures embodied in user experiences of fintech platforms, reflecting enduring practices of trust and risk-sharing \cite{kotut_terms_2025,ankrah_social_2025}; and shifting policy and regulatory contexts, organized around consumer data protection and a new class of ``public-private'' market interventions \cite{upadhyaya_digital_2025}. 

Based on extensive ethnographic fieldwork conducted between February and November 2025, our study followed the technical, organizational, and cultural practices of data science and product teams seeking to design and build new credit scoring systems for the Kenyan market, along with the practices of borrowers, recipients and communities seeking to access and manage their relation to these systems. Here, we report primarily on the practices of startups, telcos, and banks to examine how they seek to formulate and translate abstract concepts of credit risk into credit scoring tools and systems with significant market and real-world impact. 

This paper argues that algorithmic credit scoring in Kenya is accomplished through the ongoing acts of alignment that stabilize risk under conditions of persistent uncertainty. We show that risk and uncertainty are not separate phenomena but deeply entangled: what credit scoring models claim to measure as risk is persistently disrupted by three forms of uncertainty about what can be known, whether models work as claimed, and the discretion exercised over model predictions. Drawing on theories of alignment from \citet{fujimura_constructing_1987} and \citet{dourish_allure_2021} showing how technical problems are made ``do-able'' across organizational and social worlds, this paper shows how credit scoring is made to work despite the entanglement of risk and uncertainty. In contrast to the version of alignment that appears in AI safety discourse \cite{ahmed_field-building_2024,ji_ai_2025}, we argue that alignment works as a \textit{two-way} translation: making models safe for worlds, but also worlds safe for models. This double effect is central to the efficacy of credit scoring, but also its social impacts, good and bad, in the crucial and contested realms of privacy, surveillance, and financial inclusion.

We begin by reviewing HCI and related work on risk, uncertainty, and alignment, and how these concepts interact in high-stakes decision domains such as algorithmic credit scoring. We then describe our research sites and methods. Our findings document how practitioners construct alternative data using non-traditional sources of data, illustrate how risk is constructed in multiple ways, and highlight the processes through which model performances are negotiated. Our discussion argues that the making of alternative data is \textit{performative}, reproducing varied effects of risk; examines how the entanglement of risk and uncertainty exposes the limits of prediction; and discusses the important work of alignment, how it works, and for whom. We conclude with some questions for policy and sociotechnical research.

\section{Risk, uncertainty, and alignment}

The wide use of automated-decision systems in high-stakes domains has attracted significant attention and concern across HCI and human-centered ML \cite{bjorkegren_manipulation-proof_2020,lum_predict_2016,marda_data_2020,mehrabi_survey_2022,ramesh_how_2022,selbst_fairness_2019,wang_against_2024}. HCI and allied fields have studied the diverse contexts in which “risk” is formulated and operationalized in predictive systems, including credit scoring \cite{hurley_credit_2016,passi_trust_2018}, social work and welfare \cite{coston_validity_2023,kawakami_studying_2024,saxena_human-centered_2020}, criminal justice \cite{angwin_machine_2016,kasy_fairness_2021,metcalf_taking_2023}, and healthcare \cite{chancellor_who_2019, ruckenstein_datafication_2017,ismail_public_2023}. Recent studies in HCI have questioned the epistemic foundations of risk, how it is formulated and operationalized in predictive models, and the imperatives, organizational practices, and social worlds that undergird it. For example, Saxena et al.’s \cite{saxena_rethinking_2023} computational narrative analysis of child-welfare case notes in the US has found that failures in welfare cases arise from caseworker omission of mistakes in administrative data as well as asymmetrical power relations between caseworkers and parents, obscuring systemic and temporal risk factors that exceed sanctioned “static constructs” of risk. \citet{wang_against_2024} have argued that predictive optimization systems reduce complex social and organizational problems to technical optimization tasks, often amplifying the very risks they aim to mitigate. This growing line of inquiry resonates with foundational work in history and sociology of science that established that the designation of risk as calculable is not neutral, and was achieved through long and enduring processes of classification and quantification \cite{desrosieres_politics_1998,porter_rise_2020,bowker_sorting_2000}. The economist Frank H. Knight \cite[p. 233]{knight_risk_1921}, in his famous elucidation of risk and uncertainty, distinguished risk as measurable “either through calculation \textit{a priori} or from statistics of past experience,” and uncertainty as unmeasurable because it was shaped by (future) factors that could not be anticipated. Decision-making in business and adjacent domains was therefore not simply about managing quantifiable risks but rather operating in the context of an unknowable future \cite{knight_risk_1921}, making the management of uncertainty (or ``the taming of chance'' \cite{hacking_taming_1990}) a crucial site for success or failure \cite{amoore_politics_2013,appadurai_ghost_2011,carruthers_uncertainty_2013}. In these accounts, what is knowable and thus countable as risk is always in flux – shaped by evolving techniques of measurement, probabilism, sense-making, and data analysis, as well as by the contestations and negotiations of diverse stakeholders.

Credit scoring provides an important and consequential site for examining the construction of risk in algorithmic systems: both because practices in this area are undergoing rapid and significant transformation (perhaps especially in parts of the majority world like Kenya, which, as the following accounts will show, have emerged as a kind of laboratory or testing ground for new financial and algorithmic arrangements); and because such practices matter, with real and immediate effects on life chances, the ability (or not) to participate in local and global economies, and access to key resources from banking to education to healthcare. While the longue durée of credit risk spans multiple cultures, national contexts, and timeframes \cite{graeber_debt_2012}, the development of algorithmic credit scoring has been crucially influenced by historical and regulatory transformations in the US \cite{carruthers_uncertainty_2013,lauer_creditworthy_2017,poon_systemic_2010}.\footnote{Outside the US, countries like China \cite{liu_multiple_2019} and the UK \cite{aggarwal_when_2023,langely_equipping_2014} have implemented versions of credit scoring customized to their economic contexts and political landscapes. Meanwhile, the GDPR and the EU AI Act impose strict regulations on AI-driven credit scoring, emphasizing data protection, fairness, and transparency to ensure consumer rights and prevent discrimination.} Credit scoring as it arose in the late 1800s was premised on the use of personal and business information to identify and segregate borrowers based on their moral character and financial standing \cite{lauer_creditworthy_2017}. However, it was not until the 1950s, “when personal information was sufficiently standardized, computerized, and rendered objective” \cite[p. 33]{kiviat_credit_2019} that credit scoring systems “redefined creditworthiness as a function of abstract statistical risk” \cite[p. 183]{lauer_creditworthy_2017}. Fair Isaac Corporation’s FICO became the industry standard, transforming laboriously collected paper trails of personal information into three-digit statistical representations of trust and creditworthiness. Beyond transforming decision-making in lending, credit scoring models also unlocked new business models, including those related to risk-based pricing in insurance, auto loans, and mortgage lending \cite{kiviat_credit_2019}. According to \citet{lauer_creditworthy_2017}, these developments occasioned major debates over consumer protection, data bias and data privacy, producing paradigmatic regulatory interventions from the Fair Credit Reporting Act (1970), limiting information about borrowers and mandated transparency for credit denials, and the Equal Credit Opportunity Act (1974), making it illegal to deny credit based on protected attributes like race, gender, nationality, or religion and mandating explanations in credit scoring decisions. Following the growth of data mining in the 1990s, the Gramm-Leach-Bliley Act (1999) imposed new transparency rules around the transit of financial information within and between firms, including disclosure requirements around the sharing of consumer data collected in one context for use in another (a concern later theorized as contextual integrity by Nissenbaum \cite{nissenbaum_contextual_2019}).

While algorithmic credit scoring has undergone multiple stages of translation, stabilization, and standardization across industries and countries, new questions and concerns have emerged since the turn of the new millennium following technology changes and economic and political upheavals and the increased application of risk-based scoring to various aspects of life \cite{fourcade_ordinal_2024}. For example, the crises that followed 9/11 and the 2008 global financial meltdown blurred the boundaries between risk and uncertainty, and their influence is evident in the design of a wide range of predictive systems. Matters of uncertainty have become just as important as the computable categories of risk: “uncertainty masked as risk nevertheless remains uncertain” \cite{carruthers_uncertainty_2013}. At the same time, these crises gave rise to new transnational modes of data and technology governance forged out of institutional arrangements between diverse sets of state and non-state actors \cite{dourish_seeing_2007,hassan_machine_2025,singh_seeing_2021}. Political geographer Louise Amoore \cite{amoore_politics_2013} provides a compelling account of this transformation. She argues that while risk is a principal technology linking contemporary domains of security and economy, the changing formulation of risk – in response to ongoing economic crises and security demands – has adopted features more similar to uncertainty. Risk is \textit{speculative}, operating beyond probability into realms of (uncertain) possibilities \cite{amoore_politics_2013,sengers_speculation_2021}. Although everyone under this mode of data governance is framed as a subject of risk, the structuring of risk is no longer about an “individual” but rather “a fractionated subject whose risk elements divide her even within herself” \cite[p. 8]{amoore_politics_2013}. In this way, a massive data sweep is instituted to capture fractionated aspects of risk, drawing on many forms of data and also “non-scientific forms of knowledge that are intuitive, emotional, aesthetic, moral, and speculative” \cite[p. 9]{amoore_politics_2013}. Sociologists \citet[p. 83]{fourcade_ordinal_2024} frame this as a “data imperative,” where the search for economic efficiency is supplanted by the normalization of extensive data collection programs “even when [organizations] do not yet know what to do with the records they have collected.” These programs have dramatically expanded what can count as credit-risk relevant data as \textit{all} data is forcefully rendered and repurposed as credit data \cite{aitken_all_2017} -- and borrowers' lifeworlds are (re)made in data's ``image'' \cite{bowker_biodiversity_2000}. The net result of this work is a new “politics of possibility” engineered and managed by disparate groups of actors – the solutions of risk consulting firms, offerings of transnational data corporations, and interventions of international development agencies, etc. – in response to technological, economic, and political transformations \cite{amoore_politics_2013}. 

As work in HCI, CSCW, and AI ethics has emphasized \cite{dourish_allure_2021,cooper_accountability_2022,fahimi_articulation_2024}, these changes are in turn reflected in and enacted in the often backgrounded forms of work and broader contexts where they are formulated as \textit{problems} for inquiry. \citet{fujimura_constructing_1987} has argued that the \textit{do-ability} of problems depends on how problems are specified and how much they can be aligned across diverse groups of stakeholders. Alignment is only achieved by linking three levels of organization: experiment, laboratory, and social world. HCI and AI ethics scholars have adapted this framework to data science and ML \cite{dourish_allure_2021,fahimi_articulation_2024}, demonstrating how data and computing interact with analytic methods (experimental level), software tools and platforms (laboratory level), and broader institutional and funding arrangements, and legal structures (social worlds level). Accordingly, alignment is the technical and discursive work that supports coordination and collaboration across multiple levels of practice, united by commonly shared problems and interests (Cf. ``alignment'' in AI safety discourse \cite{ahmed_field-building_2024,ji_ai_2025}, understood as constraining models to follow predefined human objectives and values). 

Alignment in contexts where credit scoring is the \textit{problem} is in part shaped by advances in analytic methods and tools \cite{amershi_power_2014,mao_how_2019,muller_how_2019} which center explanatory accounts of how models function, in turn influencing varied organizational practices and normative considerations of broader social worlds. More recently, there have been concerted efforts to understand model structures and their outcomes through a range of human-centric explainability tools \cite{arya_one_2019,bellamy_ai_2018,liao_human-centered_2022}. These tools have facilitated comparisons amongst different models and outcomes, providing justification for technical designs while also meeting legal requirements. For example, Fair Isaac has claimed to adopt simpler ML regression models because of ease of explainability and comparable performance to more complex but opaque models \cite{fair_isaac_corporation_machine_2018}. In their ethnography of corporate data science teams, \citet{passi_trust_2018} studied the case of “special financing,” where borrowers with damaged credit scores or limited credit histories are matched with lenders in the US. In the process, the data science team transformed a “matching problem” into a “classification task,” in which borrowers above a certain threshold (credit score: >500) were selected for financing. Although predictive measures of accuracy “provide[d] calculative relief in the face of uncertainty,” the work of justification and explainability was ultimately \textit{collaborative} in nature, operating through organized practices of trust and skepticism, measurement and assessment, and credibility and doubt \cite{passi_trust_2018}. While the joint pursuit of explainability and performance improvements is one of the major ways through which alignment is achieved across diverse organizational and social worlds, less attention has been paid to how those worlds and the worlds beyond them are reshaped to fit model assumptions.

When predictive systems encounter the world beyond the lab, more complex forms of uncertainty appear. \textit{Modeling} uncertainty emerges from the gap between what models claim to do and the realities they face once deployed. \citet{kou_dead_2025} argue that technical claims in ML practice – e.g., economic efficiency, predictive accuracy, etc. – are often detached from real-world contexts because of ML’s optimization for industry-standard metrics (e.g., Gini coefficient), which are narrow representations of reality \cite{malik_hierarchy_2020}. The authors warn against the assumption that the gap could be easily closed with “enough societal inputs” captured in socially and culturally relevant datasets \cite{kou_dead_2025}, arguing that such an assumption only feeds into the “cultures of improvement” that pervade data science and software work \cite{bialski_middle_2024}. 

\textit{Contextual} uncertainty is a product of discretion exercised over model predictions. \citet{paakkonen_bureaucracy_2020} have equated algorithmic systems to bureaucracies, and argue that despite claims of rigidity and rationality, algorithmic systems are discretionary spaces that cannot fully anticipate real-world empirical contingencies. Although these systems can potentially \textit{replace} or \textit{enhance} human capacity, they do not do so with machine-like rationality. Thus, the focus on uncertainty allows us to “analyze how discretion works jointly with formal rules to produce decisions” \cite{paakkonen_bureaucracy_2020}. The authors add that attempts to convert various forms of uncertainty into calculable risk factors do not eliminate human discretionary power, “but rather \textit{redistribute it to work at [other] locations of uncertainty} which may go unrecognized” (italics in original text). These invisible and unrecognized spaces promote workarounds that keep systems “functional” but might also enable strategic actions that could be considered adversarial. What is often labeled strategic behavior or “gaming the system” is a constitutive feature of algorithmic systems \cite{ziewitz_rethinking_2019}, where users reshape the system’s behavior and the very definition of performance success.

Altogether, risk is not a naturally occurring measurable attribute, but it is a historically contingent construct. Although renewed interest across HCI and allied fields is grappling with the theoretical and practical limits of risk in sociotechnical systems, most of this research treats ``risk'' as a predictive optimization problem. But what counts as risk has dramatically expanded following recent political, economic, and technological shifts, which have blurred the boundary between risk and uncertainty. As such, at least three forms of uncertainty can be read from the recursive relationship between risk and uncertainty: \textit{epistemic} uncertainty arises from limited understanding of what can be known and accurately measured (per the early work of \citet{knight_risk_1921}); \textit{modeling} uncertainty reflects the limitations of predictive algorithms in capturing complex real-world dynamics \cite{kou_dead_2025}; and \textit{contextual} uncertainty shows how predictive systems are disrupted by changing spaces of human discretion \cite{paakkonen_bureaucracy_2020}. While most HCI and human-centered ML research has generated important insights on the practices of developing and deploying predictive systems in different application domains, that work overlooks the expert and pragmatic work of alignment particularly where risk and uncertainty are deeply entangled. If risk names the stabilized categories through which creditworthiness is assessed, and uncertainty names what exceeds or destabilizes such categories, then alignment is the practical, ongoing work through which stability is achieved, although alignment itself is never truly completed: it is always provisional.

\section{Histories of credit scoring in Kenya}

The history and contemporary practice of credit scoring in Kenya and East Africa have been shaped by three central forces: the long period of British colonial rule from 1895 to 1963; business transformations driven by microfinance and mobile telephony over the last 30 years; and the enduring presence of cultural and institutional forms of trust and credit that both pre- and post-date the colonial period. Echoing patterns elsewhere – for example, the enduring practice of ``redlining'' in the US \cite{rothstein_color_2018} – credit access and assessments of creditworthiness were primarily organized along racial and ethnic lines throughout the colonial period, reflecting both institutional and cultural inequities \cite{donovan_money_2024,shipton_credit_2010}; as \citet{donovan_money_2024} has noted, a 1961 World Bank report found that British banks’ preference for lending to Asian merchants at the expense of [East] Africans indicated both a social and an informational problem: “because the British bankers did not socialize with `the African population, the gulf was too wide' to accurately assess [Africans’] creditworthiness” \cite[p. 108]{donovan_money_2024}. At the same time, local governments and colonial authorities urged citizens to cultivate “saving mindsets” \cite{donovan_money_2024} while laying the foundations for documentary evidence that could be used to build and verify creditworthiness \cite{bruhwiler_moralities_2015}. This emphasis came at the expense of enduring “grass-roots” structures such as rotating savings and credit associations which leveraged culturally specific forms of promise and obligation in socio-economic relations \cite{shipton_credit_2010}. Postcolonial development interventions like microfinance have drawn on these informal logics, seeking to blur the distinctions between those who are risk-taking (and ``entrepreneurial'') and those ``at risk'' (the underserved poor) by framing poor people as capable of becoming self-sustaining microentrepreneurs through access to microcredit. \cite{roy_poverty_2010}. The Grameen Bank microfinance model, founded and evangelized by 2006 Nobel Laureate Muhammad Yunus in Bangladesh in the 1970s, heralded a new form of social lending in which microloans were distributed in “solidarity groups” often comprising about four women who acted not only as guarantors to each other but collectively ventured into small-scale entrepreneurship \cite{shipton_credit_2010,roy_poverty_2010}.\footnote{The local NGOs and welfare organizations that replicated microfinance in places like Kenya excelled at differentiating between “good women” and “risky men” \cite{roy_poverty_2010}, ensuring that loan repayments were enforced not only through gendered techniques of peer pressure but also the force of local chiefs and police officers \cite{shipton_credit_2010}. But the globalization (and indeed financialization) of microfinance came with significant costs, eroding much of the promised social impact and financial responsibility. The microfinance debt crisis that came to a head in places like Andhra Pradesh (India) in the 2010s entangled credit with despair, high default rates, and suicide or death \cite{shipton_credit_2010,roy_poverty_2010}.}

Credit regimes in Kenya underwent an important phase shift with the growth of mobile telephony and the advent of M-Pesa – a peer-to-peer money transfer technology first successfully launched in Kenya by Safaricom in 2007 \cite{guma_incompleteness_2020,mwesigwa_airtime_2024} – which fundamentally changed the structure of consumer microloans, shifting the focus from microfinance and opening paths for new and experimental models of digital lending under the development-driven banner of “financial inclusion.” The growth of contemporary algorithmic credit scoring in Kenya can thus be traced through three phases: i) the successes of M-Pesa, which provided a data infrastructure on which digital credit products could run (i.e., M-Shwari), and the institutionalization of credit information sharing regulations within bank and non-bank entities; ii); the turn to alternative data (and ML techniques) amongst app-based credit providers seeking freedom and autonomy from Safaricom’s monopoly control on data infrastructure, and iii) the codification of consumer data protection laws and the founding of the “Hustler Fund,” a government digital lending initiative launched in November 2022.

In the first phase, the launch of M-Shwari in 2012 by Safaricom and Commercial Bank of Africa (CBA, and now NCBA) heralded a new savings and loans product that leveraged one of the first well-documented experimental algorithmic credit scoring models \cite{cook_how_2015}. This product enabled Safaricom subscribers to access unsecured short-term CBA-backed digital loans between KES 100 - 50,000 (~US\$1 - \$500). While M-Shwari soared in usage, its true innovation was the extensive data infrastructure that supported it, which combined call detail record (CDR) and airtime consumption patterns in developing the product’s scorecard. This data was complemented by the Integrated Population Registration System (IPRS), the government’s automated nationwide database of citizen demographic and biometric data, and data from the mandatory SIM card registration required after 2013 \cite{breckenridge_failure_2019,donovan_rise_2014}. At the same time, the Central Bank of Kenya (CBK) instituted a credit information sharing framework that mandated banks to share credit information, eventually resulting in the licensing of three Credit Reference Bureaus (CRBs): TransUnion (2010), Metropol (2011), and Creditinfo (2015) \cite{magale_towards_2024}. These CRBs quickly accumulated negative listings (reports of defaults, etc.), including from the non-banks that shared information voluntarily \cite{breckenridge_failure_2019}. By 2020, over 2.6 million people had been negatively listed at CRBs, and about 15\% of them owed debts under US\$2 \cite{zollmann_living_2020}.

The second phase was characterized by the widespread adoption of ML techniques and alternative data, marked by the entry of Tala (2014) and Branch (2015), both Silicon Valley-based startups, along with a rush of other non-bank and non-telco upstarts in the region. These digital lenders developed app-based digital credit products powered by proprietary credit scoring models, which operated outside of closed banking or telecommunications data infrastructures and their associated regulatory requirements \cite{upadhyaya_digital_2025}. While these new entrants afforded new possibilities of credit access for some, they also raised major concerns around consumer privacy breaches, aggressive debt collection, and predatory lending practices \cite{munyendo_desperate_2022}. As summarized by digital lending scholars \cite{donovan_perpetual_2019} “[these digital credit apps] give you money gently, and then they come for your neck.”

In the third phase, the Kenya Data Protection Act (2019) was enacted, alongside the Data Protection (General) Regulations (2021), which stipulated \textit{privacy by design} and \textit{privacy by default} principles (although subsequent studies have found continued gaps in how consumer data is shared with third parties \cite{mutungu_privacy_2021}). The Act also established the Office of the Data Protection Commissioner to receive complaints from the public; relatedly, the Central Bank’s continued solicitation of public comments on a raft of regulations represents efforts to legitimize and safeguard digital lending \cite{upadhyaya_digital_2025}. Meanwhile, in 2022, the newly elected Ruto government announced its “Hustler Fund,” an initiative aimed at providing digital loans to underserved youth and MSMEs (the so-called “hustlers”) \cite{thieme_hustle_2025}. While the Fund is supported by a consortium of banks and telcos and has widely publicized its use of credit scoring, it has also recorded the highest default rates in the country, with a non-performing loan (NPL) rate exceeding 68\% \cite{kenya_human_rights_commission_failing_2025}.

While each of these phases revealed distinct logics of credit scoring, their common theme is around the use of data to demarcate risk associated with particular borrowers and groups. This subtle move disembeds the individual borrower from their broader social networks, while defining disciplinary actions for non-compliance, ranging from sanctioned negative listings to threats and harassment in debt recoveries. Yet as \citet{shipton_credit_2010} has shown, certain cultural forms and local practices around wealth, credit, and obligation endure across political and technological transformations, persisting as contexts that algorithmic systems must accommodate but cannot fully anticipate. In the sections below, we will present concrete and real efforts by data science and product teams in grappling with these situated cultural practices, drawing on large datasets and expanded definitions and evaluations of risk.

\section{Research sites and methods}

\begin{table*}[htbp]
\Description{A table displaying 30 interview participants organized into five domain categories: AI \& Data (13 participants), Financial Services (8 participants), Consulting (2 participants), Policy \& Regulation (4 participants), and Development \& Investment (3 participants). The sample comprises 21 men and 9 women, with gender indicated by [M] for men and [W] for women next to each participant ID. Two participants [P10] and [P30] were each interviewed twice.}
\centering
\renewcommand{\arraystretch}{1.4}
\begin{tabular}{@{} l p{12cm} @{}}
\toprule
\textbf{Domain} & \textbf{Participant IDs} \\
\midrule
AI \& Data & P8 [M], P9 [M], P10 [M]$^{**}$, P11 [W], P12 [M], P16 [W], P17 [M], P20 [M], P22 [W], P26 [M], P27 [W], P29 [M], P30 [M]$^{**}$ \\
\addlinespace
Financial Services & P1 [M], P2 [W], P3 [M], P4 [M], P13 [M], P18 [M], P19 [M], P23 [M] \\
\addlinespace
Consulting & P21 [M], P24 [M] \\
\addlinespace
Policy \& Regulation & P6 [M], P14 [W], P15 [W], P25 [W] \\
\addlinespace
Development \& Investment & P5 [W], P7 [M], P28 [M] \\


\bottomrule
\multicolumn{2}{@{}l}{\footnotesize\textit{Note:} [M] = Man; [W] = Woman; [ ]$^{**}$ = Two interviews} \\
\end{tabular}

\caption[Interview Participants]{Overview of interview participants drawn from five broad domains. The sample includes 21 men and 9 women (N=30)}
\label{tab:participants}

\end{table*}

This paper builds on nine months of ethnographic fieldwork in the AI and algorithmic credit scoring worlds of Nairobi, Kenya. Between February and November 2025, the first author, who is from Uganda, lived in Nairobi, where he examined the sociotechnical and institutional practices of four startups and various financial and policy actors involved in designing alternative credit scoring systems. Nairobi was selected as the principal “field site” because of its global successes in financial innovation and because of the first author’s knowledge of its tech and startup ecosystem, which he has followed and participated in for over a decade as a tech journalist. Although the study was led and conducted by the first author, who was known to his interlocutors as a social scientist of AI and credit scoring based on his professional and academic standing, he was supported and advised by a team of co-authors who are researchers from North America with established histories of research and collaboration around technology and development projects across various parts of Africa. 

Participant observations were conducted at startup and bank offices, technology events and policy workshops, and through online channels, including LinkedIn, WhatsApp, and Twitter/X \cite{de_who_2025}. The first author had been contacted in February 2025 via LinkedIn by a senior researcher at a startup, Moonshot,\footnote{Moonshot and all names of startups, products, and participants used in this paper are pseudonyms.} who expressed deep interest in his prior work on mobile money in East Africa \cite{mwesigwa_airtime_2024,csikszentmihalyi_space_2018}. He explained that human-centered insights are important to Moonshot, a remote-first company that develops proprietary algorithmic risk and credit assessment products for both regulated and unregulated lenders and also runs an app-based financial marketplace that connects individuals to digital lenders for loans (while also receiving personalized financial health advice from the app). The other startups observed were remote-first Kenyan-based companies building related but differentiated AI-driven credit scoring products: Startup A was bundling loan management systems and algorithmic credit scoring solutions exclusively for unregulated financial institutions (such as savings and credit cooperatives, which are very popular in Kenya); Startup B offered technical systems to overcome compute and inference limitations for tasks like on-device credit scoring; and Startup C was developing lending data infrastructure for solutions such as credit scoring. Moonshot was the largest and most prominent, with a team of 10 to 20, while the rest all had fewer than 10 staff. Most of the teams worked remotely, with occasional visits to different co-working spaces in Nairobi or field visits to their clients, including banks and non-bank digital lenders.

Through informed consent, the first author was granted partial access to conduct in-person participant observation during startup work sprints and company events, as well as at banks that startups were cultivating relationships with. Reflecting the distributed and virtual nature of the startup teams and worksites, these observations were sporadic and often informal, held whenever the startup teams were in the city (every other month). Our interest was in how people building these AI systems collaborated and thought together about their work, including through practices of data collection and risk formulation, and how these models were evaluated across multiple technical benchmarks and business goals. As we disclosed our research interests, other interlocutors, including data scientists, business people, and policy analysts were engaged across online and offline channels, where social and intellectual discussions about credit scoring and adjacent topics were held. Given the frequently fluid and network-based nature of startups in Nairobi (and indeed other tech capitals), these methods followed a tradition of ``milieux'' or ``scene'' analysis that has been adopted by ethnographers of high-tech work from \citet{saxenian_regional_1996} to \citet{neff_venture_2012} to \citet{shestakofsky_behind_2024}. While the worlds and practices observed could reflect the Silicon Valley imaginary of “digital entrepreneurship,” it would be a mistake to assume that the work and enactments of our interlocutors were simply replication or mimicry: local actors created and adapted tools and practices that have since circulated globally \cite{friederici_digital_2020}, and in some cases at least the experimental practices disclosed went beyond practices known to be common in the better-regulated credit markets of Europe and North America. Indeed, some of the firms present, including Europe- or US-based, appeared to be attracted by the prospect of a space of less trammeled, less regulated experimentation which could produce lessons and innovations that could then be brought back to North America and Europe -- a kind of flow or directionality (“ICTD in reverse”?) that HCI and computing fields have been arguably slow to recognize (though anthropologists \cite{comaroff_theory_2012} have been quicker to observe). In these fragmented and distributed worlds of credit scoring practice, gossip and secrecy mutually co-existed \cite{marcus_ethnography_1995}: for example, some actors wanted to know what the first author knew (presumably about their competitors in the credit scoring market) or what exactly he wanted to find out, while also equivocating about what exactly their work entailed (wrapped under NDAs and tightly kept company secrets). Altogether, several participants shared and interacted formally and informally across events, interviews, and online correspondence - though one prospective respondent declined an interview request.

The first author documented and transcribed over 150 pages of field notes and transcripts from observations with the startups and other interlocutors noted above, as well as 11 technology events and archives of over 300 web links, screenshots, and images. The events and workshops attended varied in focus and size, from the small and local to Kenya’s premium annual startup and tech expo with over 3,000 participants from Kenya, Nigeria, Uganda, the UK, and beyond. Numerous informal (unrecorded) discussions and observations in co-working spaces and other settings were also held and captured in regular field notes produced by the first author and shared and discussed within the author team. In addition, an anonymized dataset with over 80,000 observations (rows) and more than 260 features (columns) on Kenya-specific credit scoring, along with a private report and pitch deck, was supplied by one of the startups.

In addition, 32 in-depth interviews (totaling roughly 38 hours of recorded content) were conducted either in person (8) or on Microsoft Teams, Zoom, or Google Meet (24) with 30 participants (see Table~\ref{tab:participants} for the breakdown). While the first author is fluent in English and Swahili, and some conversations and exchanges occurred in and sometimes across these divides, formal interviews were conducted in English, an official language of Kenya, commonly spoken in professional and informal settings. The transcriptions from field notes and interviews were revised and edited for accuracy and completeness, and subsequently shared and coded by the first author in consultation with the research team following principles of Grounded Theory in HCI \cite{muller_grounded_2010}. Through analytic comparisons between field data and key moments in the field (including tech events and a chance Twitter/X Space), thematic coding was used to develop key code groups such as “risk,” “credit cultures,” “narrative,” and “fintech actors,” together with the themes of risk, uncertainty, and alignment developed in the paper.  The sections below present findings from this fieldwork, with particular emphasis on the making of alternative data, the multiple constructions of risk, and the negotiation of model performance. Each of the three scenes opens with an ethnographic field note (in \textit{\textbf{italics}}), followed by additional insights and analyses grounded in subsequent interviews and observations. 
\section{Findings}

\subsection{SCENE ONE:  Making alternative data}

\textit{It’s a gray morning. It’s been raining heavily since the early hours of the day. Under a well-lit and air-conditioned conference hall in an upscale part of Nairobi is an ``open data'' for finance workshop. There are about 50 participants in the hall, surrounded by a camera crew. Live feeds from the workshop’s proceedings are broadcasting on a Jumbotron-style screen at the main stage. The event is kicking off in earnest. The moderator calls the organizers to the main stage. Representatives from the organizing consortia of industry, players in the prudentially regulated banking sector, an umbrella of non-deposit-taking [digital] lenders whose regulations are being reworked by the CBK, and a development-focused, financial policy-influencing body join the moderator on stage, standing in a line and joining arms with each other on either side. The moderator says that these institutions want to work together, sharing and leveraging each other’s data, adding that the first and last time such a workshop took place in Kenya was in 2018. It’s 2025, and the banks, telcos, fintechs, and credit reference bureaus (CRBs) want to try again. A facilitator is called to take the stage, and he announces the agenda for the day: open data for finance.}

{\em As a slide presentation rolls on the giant screen, the facilitator outlines the key topics. “Open data is about the consumer at the center,” he declares, “it’s a consumer-led data-ship.” He shows successful examples of the open data model around the world, leading with an example from Brazil that has 68 million active accounts generating data that is accessed weekly through “2 billion+ API calls.” He closes with a question, asking participants: ``what would it take for open data to succeed in Kenya?” The audience chimes in, one person after another, all calling out Safaricom.\footnote{  Although Safaricom is the leading telco in Kenya (with 50 million subscribers out of a country of 56 million people), it operates in a legal gray area. It’s not a prudentially regulated financial entity, yet M-Pesa and its suite of financial products dominate 95\% of the digital credit market. Beyond fintech, Safaricom is central to Kenya’s security and bureaucratic apparatus, and imaginations of progress and innovation.} They say that Safaricom has evolved into a national payments data chokepoint, and open data for finance can only succeed with the end of Safaricom’s monopoly. A Safaricom representative defends the company, arguing that Safaricom has enabled competition in the fintech space and broader banking sector. A few heads nod in agreement, but the salvo of questions continues. The moderator chimes in to suggest that maybe Safaricom shouldn’t be talked about anymore. The audience breaks into laughter. In the next session, a facilitator reminds the audience that not all data is the same, laying out three types of data for open data: KYC (know your customer), customer financial data, and market insights. While the CRBs are mandated to collect and share credit information data, quality and coverage issues abound. Representatives from telcos and banks are passionately arguing against sharing ‘their’ data - that they have spent significant resources acquiring data and that they wouldn’t share under the proposed open data framework, despite the framework’s emphasis on consumer privacy. As the workshop ends, it seems that much is left unsaid; a person from the startup that secured my invitation to the workshop reflected that the workshop was “generic” and “ordinary.” A representative from another startup whom I shared a meal with boasts that they have amassed more data on MSMEs than the CRBs.}

The field note paints a picture of the stalemate that has characterized Kenya’s financial data landscape for more than a decade, where industry efforts to share data have not realized sufficient momentum to benefit all players. While CRBs have built expansive datasets accessible to all (for a fee), Safaricom’s monopoly of a vastly superior real-time data infrastructure is exclusionary. And yet, upstarts are seeking alternative means to access Safaricom data (and from other sources), and are seemingly making progress. Below, we show how a group of actors is leveraging technical means to collect and analyze massive volumes of data, and also highlight the fragility of alternative data infrastructures.

\subsubsection{Data collection and classification}
Several of the startups we spoke with reported technical workarounds they were deploying to gather real-time data about financial transactions, particularly for those considered “thin-file” or “credit invisible” (those whose credit information is not captured in the CRBs; most belong to demographics such as the youth and rural women). Through these methods, data collection could be automated, scraping data directly from users’ devices or via APIs linked to third-party data platforms. Amongst these, \textit{“SMS scraping”} was seen as the gold standard by many of the participants in our study. Access to user SMS/text messages provided significant volumes of transactional and behavioral information about a phone user, covering transactions from utilities to food to health. According to P11, the CEO of Startup C, SMS logs were an important source of accessible and verifiable personal data:

\begin{quote}
    \textit{“So we said to ourselves, where did the highest form of financial records sit? The highest form of financial records in emerging markets [like Kenya] actually sits in SMSes. OK, the highest form of financial records in the US is an e-mail. Every time you go and buy something in the US, you receive an email confirmation. So if I was to ever build the same infrastructure in the US, I would do an e-mail parsing machine learning infrastructure because that is the highest form of financial records, everything you pay for is recorded there…the sweet spot in emerging markets and places like Kenya is we have mobile money [e.g., M-Pesa], which is telco-based and holds so much of the data.” } \\
    -- \textsf{[P11], CEO of Startup C}
\end{quote}

Startups reconstructed M-Pesa statements in real-time by scraping and parsing SMS data, mining the message alerts from M-Pesa, banks, and other providers (alerts sent to users after every transaction). Although the data from the workarounds could sometimes be partial, once paired with other data, they became sufficient in generating useful insights. 

\begin{quote}
    \textit{“Through again working with the fintechs, banks have also made some advances to try and see how they can collect this information and I’m sure probably you have had some apps that are able to snoop through your phone [through software development kit (SDK)]. Of course, you have to do the right permissioning on the app so that you allow them to snoop the data. So banks have worked with fintechs to get some apps that are able to snoop data from your SIM card to see your M-Pesa statements, your airtime purchases, other banking alerts, and we try to build back the same information that Safaricom would have but of course that is not 100\% accurate. But it’s giving banks an opportunity also to get a share of the pie.”} \\
    -- \textsf{[P3], Product lead in a bank, with decades of experience in banking}
\end{quote}

The technical expertise in developing SMS parsers for data collection in turn relied upon cultural and situated knowledge to interpret, annotate, and classify the data collected. For instance, while Moonshot had used their suite of SDKs to collect over 500 million data points from user devices, the team of CTO and junior data scientists invested time to interpret and classify the data collected. P10 said that he once spent four hours in a corner of a co-working space manually reading data associated with financial statement transactions, and trying to understand what was going on: \textit{“How do you extract this information from it? [...] How do people spend their money? How do people earn their money? Do they bet? Do they save?”}. It was also not unusual for a company to contract another to support the interpretive work of annotation and classification. For example, P8 provided data annotation services to a well-funded startup from Silicon Valley that wanted to make inroads into the Kenyan digital lending market. P8’s team was supplied with a dataset of 50,000 SMSes in a text-readable format, with clear instructions to identify particular entities and categories. Besides generic entities such as “TIME”, “FIRST NAME”, “LAST NAME”, “AMOUNT,” and “UTILITIES,” [P8] drew on their local experience to suggest additional entities, including “SPORTS BETTING” and “FOOD.” 

Once the data pipelines were completed (and automated), our interlocutors suggested that they could collect and analyze data in multiple ways. Customer data profiles were available on demand, and comprehensive reports could be generated in \textit{“20 minutes, real time,”} according to P11, who added that, \textit{“customer data profiles have everything: their affordability, the loans they have, do they gamble, where they eat [...] For us, credit is the thing that you got to do [to make ends meet]. There’s so much this data can do, right? You know, down to even healthcare.”}

\subsubsection{Regulatory demands and technical dependencies}

Our interlocutors spoke about industry pressures to diversify their data, including through formal and informal data sharing partnerships with actors such as government agencies (for example, agriculture data or satellite data) or even sports betting companies (which have amassed outstanding volumes of transactional data). An unpublished report on “Alternative Data” in Kenya commissioned by the World Bank and IFC, and circulating in draft form by July 2025 spotlighted the lack of access to credit as a major impediment to consumption and MSME (Micro, small and medium-sized enterprise) growth. The report showed that only 40\% of MSMEs are represented in traditional CRB data, and could dramatically increase to 75\% if alternative data was used. Through “consumer-led, consent-based sharing,” the report suggested that new developments in ML are enabling “this data to be extracted, transformed, and safely shared with third parties at relatively low cost.” Since the law on credit information sharing was not very clear on the legality of alternative data, interlocutors suggested that these efforts to collect additional data exerted pressure on highly regulated sectors (such as CRBs, who the World Bank had recommended to double down on alternative data). But there was still room to maneuver (and even grow). According to P12, a CEO of a data exchange platform in Kenya, there was need for more diverse sources of data, including “real-time digital payments,” and “Hustler Fund loan details.”

\begin{quote}
    \textit{“The credit bureau is good in terms of tracking payments on [debt] facilities taken, but it’s not the be all and end all for expanding the reach or growing the access to credit because there’s a lot that is actually missing that is not captured in the bureau [...] So there’s good opportunity for [CRBs] to grow banks balance sheets 2 or 3 times; whatever it would be based on this alternative data. The balance sheets are pretty thin because of what is visible, so there’s good opportunity for growth.”} \\
    -- \textsf{[P12], CEO of Data exchange platform}
\end{quote}

Yet these diverse sources of data relied upon other sources of static data, including personal information (date of birth, gender, nationality) and financial data (SIM card number and bank account). Static sources of data provided the much-needed KYC information, which could be used to verify personal or financial identities. A respondent called out the dependence of some tier-1 banks on Safaricom’s fragmented infrastructure. P19, who has worked for a large telco, said that \textit{“[within] Safaricom [...] there is no one source of truth for customer data or what we call master data.”} This meant data is pulled from multiple internal databases in a less structured manner. These fragmented infrastructures introduced additional problems downstream: porous financial identities because SIM card numbers are fungible (one’s number can be resold to another person after a definite period of inactivity). P23, who has also worked as an engineering platform manager at a large telco, said:

\begin{quote}
    \textit{“Most of the credit lenders use [SIM card] phone numbers as the basis of [sourcing and verifying borrower identities for lending]. If a phone number [...] has had 6-7 identities, and 80\% of the identities have been using [a digital loan] facility, sometimes it does present as a faulty [data] parameter to measure because the life cycle management, especially on the lender side, is often not managed.”}
\end{quote}

There are three takeaways from this section. First, the search for alternative data is almost endless, enlisting various techniques and cultural expertise. Any record that can be collected and classified will be collected and processed. Second, alternative data dramatically expands what counts as a record, especially for those who might have been overlooked by traditional data infrastructures. It is through this expansive gaze that alternative data enables different modes of differentiation. Third, alternative data can break at the seams. They can be fragile, as emergent or enduring technical, legal, and institutional issues undermine their effectiveness.

\subsection{SCENE TWO – Dissecting risk}

{\em The week is coming to an end. White sparks billow from four cold fusion machines placed around a small stage in the corner of an open-roof venue in an upscale Nairobi events space. The space is filled with middle-aged men along with a handful of women and younger people. Most are dressed in business casual, although a handful of the men are wearing suits with neckties. Some faces look familiar (to me), having met some people at different tech events in town. Three innovators stand on the stage with wide smiles, as confetti pops behind them and a recently released afrobeats song serenades the air. It’s a big day – an invite-only event for folks in the digital lending space (notably banks, telcos, and asset financiers) who are gathered to witness the launch of a new credit score: After months of rumor and anticipation, Extenscore® has landed. The score has been developed by Moonshot in partnership with an established data exchange platform in Kenya. This collaboration has been going on for over a year, with intensive efforts around research and experimentation to get the score right, the product lead announces. The lead adds that he sees close connections between how this new score will support “community,” redolent of his previous experience in banking abroad. He says that the banks he worked for extended credit in a safe way to “Gypsy” and “illegal immigrants,” and people who didn’t have permanent addresses, by using community links like churches and thereby generating trails of data about these people. He was surprised by how much lives could be changed through the “importance of credit.” It’s game-changing, he says, that something as intangible as data could have concrete outcomes, enabling acquisition of “devices” and “vans.” As the talk continues, the passion and the superlatives increase: revolutionary, innovative, record-breaking. He invites the CTO to take the stage, whom he adulates alongside the team for being “very technical.”}

{\em The CTO walks to the stage to a warm round of applause. The new score, he explains, combines financial data provided by a leading credit reference bureau with Moonshot’s alternative data acquired from sources including borrowers’ phone metadata and SMS logs. He announces that the model has been trained on a dataset of over 10 million anonymized transactions, and outperforms three industry-standard scores by orders of magnitude. These comparison scores are not named, but the statistics are broadcast on the large pixelated screen to the right of the small stage for all to see. The CTO interprets them. The score’s performance is 80\% better than the competition in selecting between good and bad borrowers, he says. The score is inclusive too, catering to MSMEs, the youth, and those in rural areas who have long been underrepresented in traditional financial data from credit bureaus. Later, the festivities give way to networking and informal conversations: “I wish I had this score when I was a loan officer,” “I built a similar decisioning engine in London,” “Machine learning models are now more explainable these days.” Such stories, shared among strangers and friends alike, circulated as the CTO moved through the room, inviting guests to connect and watch for future products from Moonshot.}

In the field note, a new credit risk score is debuted with fanfare. According to its developers, it was through their mastery of risk and the high-level techniques of data science and ML that they achieved such impressive performance results – at least on paper. Promises to revolutionize digital lending for risky subjects are discursively accomplished through specter and performance, and legitimated through partnerships with more established industry actors. In the following subsections, we show how \textit{“risk”} is constructed, including the various assumptions and hypotheses that must be continuously tested to support its legitimacy and ongoing operation. 

\subsubsection{How expert guessing prefigures feature engineering}

Data is the foundation upon which credit scoring models are built. As we have shown above, a substantial amount of work is required to collect and analyze data. However, the model assumptions that these data would support in turn depended on highly structured sets of data, used to extract features (attributes of variables) and collect labels (decisions about past loan performance). While feature extraction and engineering were technical accomplishments, initial data on labels were often acquired from CRBs, whose data coverage was poor. Therefore, most actors had to start from scratch, albeit with varying degrees of access to resources. Our interlocutors, reflecting on the initial stages of developing their credit models, described the process as \textit{“hypothesis testing”} and \textit{“expert guessing.”} As said by a participant who worked for a large bank in the very first phase of credit scoring and now works as a credit scoring consultant across several markets in Africa:

\begin{quote}
    \textit{“You need example data [to build a model] and the main issue [at the start] was there was no example data because you are yet to launch the service. So how do you start? So you have to start with some hypothesis, and one way to do this in credit scoring is something called expert scorecards. So you can do expert scorecards with hypotheses which you can take as an uneducated guess [...] Honestly, it’s guesswork, but expert guesswork.”} \\
    -- \textsf{[P30], Data scientist and consultant}
\end{quote}

Expert scorecards were more common in banks (but startups used versions of these scorecards at the start), where experienced and long-serving credit officers and risk analysts developed rubrics with arbitrary scores associated with unique cultural and behavioral characteristics.

It was only after the accumulation of data on loan performance that feature engineering enabled our actors to \textit{“splice and dice”} [P30] variables of borrower behavior into several \textit{sub}variables. Data scientists and engineers across our four startups described to us how features were extracted from a number of variables including income inflows and outflows, and then associated with local and culturally relevant behaviors. Although one could come up with hundreds of features depending on their preferences (for example, disaggregating income by time: \textit{“last month’s income, two months ago, income three months ago, etc.”}) the goal was always to have a manageable amount of features, usually below 20: companies that boasted about having thousands of features were performing a \textit{“PR exercise,”} [P10]. However the art of selecting what features to use in a model was an exercise in testing for correlations and strengths between features, a process that [P9] likened to a set of \textit{“cool tricks”} that relied on \textit{“non-scalable manual checks.”} This stage of feature engineering was as much about creativity as it was about statistical rigor, requiring constant iteration to ensure that the most predictive features were captured.

\begin{quote}
    \textit{“So for banking we normally do logistic regression with binning, just simple. So we have software tools that we use, mainly the opt binning. Then you do your iterations, do iterations until you’re satisfied with the features [...], then you fit [the model].”} \\
    -- \textsf{[P26], Data scientist at a Tier 1 bank}
\end{quote}

\subsubsection{How application and behavioral scoring rely on expert guessing}

Several participants classified credit scoring models into two broad categories: application and behavioral models. Application models assess the likelihood that a (new) borrower will pay back, while behavioral models draw on an already established credit relationship to predict loan repayment amongst existing customers. Several commercial banks and budding digital lenders relied heavily on CRB scores to generate application scorecards, despite concerns about CRB data quality, coverage, and completeness. While participants in our study spent significant time working through correlations between features and labels in their models, these models were truly tested only when they were deployed in real-world contexts. In an early experiment underwritten by an emerging digital lender who was keen to leverage Moonshot’s alternative data and ML-scoring technology, Moonshot was shocked by the performance of their initial application scoring model.

\begin{quote}
    \textit{“The very first model we did was very simple. It had total flows in the last six months, and like total flows and maybe like standard deviation of outflows, or something like that. The repayment rate was terrible, it was terrible, like it must have been something like 20\%. At the time, we’re just running experiments, so it wasn’t the end of the world, but it was pretty much like, oh, damn. So, actually, we thought that someone who has very high cash flow activity, like in amounts of 1.5 million Kenyan shillings, is going to pay for sure, especially because we are starting with small lines [loan size of KES 500]. Obviously, the higher the income, the lower the risk. But actually, it’s very often not even the main characteristic. High income is not the most predictive feature. Actually, people who have high cash flow activity don’t really mind all those small penalties you’re going to give them, so they’re going to repay when they feel like. They are taking loans with five, six different lenders, and they’re going to prioritize who they’re going to pay first. There’s just so much psychology behind it.”} \\
    -- \textsf{[P10], Moonshot}
\end{quote}

Moonshot’s challenges were not uncommon. Different participants shared that they had also learned that making credit decisions (grant or deny loan) had to be embedded in understanding of risk and behavior better. Moonshot on its part drew more from user surveys and focus groups to understand borrower psychology, particularly in application (acquisition) and behavioral (retention) models. One of the interventions was to eschew binary classification approaches (\textit{will default vs. won’t default}) altogether, and instead assign locally optimal credit amounts. Credit was granted on a sliding scale, customized for each borrower's risk profile.

\begin{quote}
    \textit{“So we have two people who we feel could each get 20,000. They have similar behaviors, and so on, except maybe some past [usage patterns] tell us something that maybe we need to be careful [about]. So, we believe the two of them have the capacity to repay [but] we feel like [one] may not do it. [So] we basically re-balance the maximum line they could get [...], which I guess in a way decreases the exposure of the lender because now, for someone who has more exposure, who has more risk, we start by giving them a smaller line and then, over time, you know, if we see that the behavior gets better, they will then get back to the 20,000 [...] If I discount your limit, I’m not penalizing you, I’m just limiting the affordability you have.”} \\
    -- \textsf{[P10], Moonshot}
\end{quote}

Yet assigning credit limits introduced, according to our interlocutors, new \textit{“risks.”} The fear that borrowers would attempt to \textit{“game your system”} and climb the credit limit ladder was framed as an ever-present \textit{“adversarial problem” }[P10]. Managing this problem relied on cultural and behavioral assumptions of fraudulent behavior, and attempts were made to avert such practices.

\begin{quote}
    \textit{So if a user comes to the [digital loan] app and tries to apply [for a loan], and that device has been used for 20 [loan] applications in the last 10 days, you know, with four different phone numbers, this is weird. So, we’re really looking at your [broader] digital footprint beyond what was reported in the credit bureau.”} \\
    -- \textsf{[P10], Moonshot}
\end{quote}

\begin{quote}
    \textit{“We discovered that if we use transactions, and the customer understands if I have a lot of credit in my account, if I use my account often, then [the limit] go up. So it’s subject to manipulation. So to prevent that, they introduced other options under parameters. There’s a feature selection model, whereby, even if you game as many times as possible, your limits cannot go beyond a certain percentage, because you’re subjected to other features [...] We use 85\% of CRB and 15\% of internal score to get your final score. We also check your affordability for the account. If the affordability of this guy is 1,500, we can confidently advance a loan of 1,000.”} \\
    -- \textsf{[P1], Data scientist at Tier 1 bank}
\end{quote}

The feedback loop between attempts to limit \textit{“adversarial behavior”} was confronted with fundamental realities of data and models: drift. In data science and ML, data and models decay over time due to shifts in user behavior and consequent loss of model predictive power. Moonshot had at least 15 models, with the majority being behavioral models catering to products such as: \textit{“Very short and 7-day, you have 28-day loans, you have three months, six months, a year, two years. Then you have car loans, logbook loans, asset financing, [etc.]”} [P10]. A Tier 1 bank had an inventory of at least 25 models serving various product segments: \textit{“We have the consumer, we have SMEs, including SME individual and SME business [...] So each product has its own scorecard from then on”} [26]. Most of our participants regularly updated their models to catch these drifts. An actor at a Tier 1 bank said that they updated their catalog of behavioral models once a month for all loan categories [P26]. Moonshot’s model updates were less regular, with [P10] arguing that instead of regular model updates they focused on fundamental assumptions around borrower \textit{“psychology,”} which instilled confidence in their data and models, and less on the force of constant updates. 

There are two key insights from this section. First, risk is formulated through expert guessing and situated expertise that \textit{“splice and dice”} data to produce multiple interpretations of risk. Second, as practitioners design for anticipated borrower behavior, ordinary practices get reclassified as suspicious or adversarial, vastly expanding the grounds for consumer surveillance.


\subsection{SCENE THREE – Balancing metrics}

{\em A member of the Kenyan government is speaking on a Twitter/X Space where users host public radio-style talk shows for free. Two social media personalities are hosting Susan Auma Mang’eni, the Principal Secretary (PS) of the State Department of MSME Development. She’s providing “conceptual clarity” about the Hustler Fund, which “is a policy intervention to address the market failures in credit access.” For the next 34 minutes, the PS speaks continuously, outlining the Fund’s purpose and emphasizies the gaps in the market it is meant to address. She speaks about the unbanked majority in Kenya, noting that the country has almost 17 million people in the informal sector, and only 3.5 million are formally employed. Those in the informal sector, she says, are considered high risk by traditional banks and are often excluded from financial services. She describes the economic situation of these “hustlers” as akin to living in a “lottery setup; today they get something, tomorrow they don’t.” The PS continues, explaining that many people in this sector rely on digital lenders charging interest rates as high as 10\% per day. This is “digital slavery!” she declares. She also however castigates Kenyans for being “serial defaulters,” and notes that when defaulters are negatively listed in the CRBs, their records remain for seven years, further freezing their access to credit.}

{\em The Hustler Fund, she says, aims to address these “market failures.” She highlights that *254\# is the USSD code to access the Fund on mobile phones. Sounding gleeful, she reminds the listeners that that is also Kenya’s country code. The digital-only Fund is accessible for free across all major digital channels, bankrolled by the government and administered by a consortium of telcos and banks. She mentions the telcos by name: Airtel, Telkom, and “even Safaricom.” The Fund uses data from the consortium to develop credit scores for borrowers. With the power of “data,” she advises, “your behavior is your property.” This, she emphasizes, is key to “widening the scope for…risk-based financing.” The lowest credit limit for new borrowers is KES 500 [about US\$4] for a 14-day repayment period. She notes that while this may seem like a small amount, she defends it by arguing that Kenya is a “kadogo economy” [small-sized economy], where 98\% of businesses are MSMEs. She adds that this figure is consistent with figures from other digital lenders and that the Fund is merely following market trends. The Fund now has the largest database about borrowers in the country. While the Fund currently has 26 million borrowers, her focus is on the 5 million borrowers who are making payments “really well.” When the host interrupts the PS, asking her to tell the story in a “human form,” she shifts tone: “the whole of this week, we’re receiving messages…please do not disband this fund.”}

The field note above captures a chance encounter on Twitter/X; in a Space that lasted two hours and seven minutes. The PS spoke for more than 90\% of the time, often uninterrupted, detailing the rationale of the Fund and how much it was excelling since its launch in November 2022. The space was attended by about 75 people, and it had only been advertised earlier on the day it was broadcast. However, earlier that week, a renowned civil rights group had published a scathing report about the Hustler Fund \cite{kenya_human_rights_commission_failing_2025}, castigating it for “failing the hustlers.” The report detailed that default rates were as high as 68.3\%, and called for the immediate “scrapping” of the Fund, citing regulatory gaps and political compromise. In the stories below, we depart from the Space’s policy-inspired monologue to show how our participants engaged with the function and role of metrics in measuring the performance of credit scoring in digital lending, and sites where practitioner discretion appear. 

\subsubsection{When failure is justified} 

According to our interlocutors, including a couple directly involved, the Hustler Fund is an example of what happens when politics enters the rule-based process of algorithmic credit scoring. As P23, an engineering platform manager at one of the leading telcos in Kenya, explained:
 
\begin{quote}
    \textit{“M-Pesa was at the core of developing Hustler Fund. So [M-Pesa] did work with partner banks and of course Airtel money was also involved and Telkom’s T-cash, but at the very core was [M-Pesa], which is like the national payment switch [...] So the challenge as had been presented to [M-Pesa] was that the credit scoring that was being done by already existing solutions was excluding the majority of the population [...] Established credit lenders [lend] very low amounts [...] There are cases where somebody qualified for only 100 shillings. And so then that meant that the [new] scoring model was going to be shifting away from what the existing lenders were doing. Gambling [aka sports betting] is a big subject as far as credit scoring is concerned. The turnover on a customer on their account is also a big item. So all this was then sort of put to the side.”}
\end{quote}

The performance of the Fund, however, has been very dismal. P23 emphasized that while the industry repayment rates for mobile loans are around the 60\% mark, Hustler Fund’s repayment rates have been “well below 40\%.” There have been various reasons why this has been the case, including the removal of important features such as sports betting. Notable among them is the elimination of negative listings in case of default: \textit{“There was no listing of the [Hustler Fund] defaulters with CRBs. That was actually communicated by the president himself.”} Other actors who developed credit scoring models suggested that including given features was a matter of developer preference. Although sports betting (gambling) is considered a negative signal, a large chunk of people including the youth partake in it. The assumption that sports betting correlates with delinquency is not unfounded, since those who participate most in it are more likely to default. However, some actors suggested that they have found ways around it: [P10] hinted that Moonshot tracked another important feature (which they asked not to be shared for proprietary reasons), whose inverse relationship allowed them to account for sports betting (without distorting their results due to negative signals associated with gambling).

P23 conceded, though, that while parameters like loan repayment would play a role in the behavioral scoring of the borrowers, from the very beginning, the business people [in the consortium] saw Hustler \textit{“as a political promise, not as a business decision, and it was acted on as such. So yes, for the businesses they might have made money, but in as far as the whole Hustler fund itself being a business case that would actually be sustainable in the long term, I don’t think that it was considered as so.” }

\subsubsection{When technical assessments encounter discretion}

Underneath the often publicly referenced NPL rates (the metric), there was a suite of technical data science and ML metrics, including the Gini coefficient that informed performance assessments of credit models under development and in deployment. Participants across data science and data-adjacent participants (especially in product) consistently pointed to the Gini coefficient, used here as a statistical measure of the differential success in predicting defaults between different models.\footnote{  The Gini coefficient is also used in economics, where it quantifies income or wealth inequality within a population, ranging from 0 (perfect equality) to 1 (maximum inequality).} The Gini coefficient measures a model’s predictive power between a repayment (good outcome) and default (bad outcome) with values between zero and one,  higher values reflecting better performance. Acceptable ranges for Gini coefficients vary by model category: as P26 explained, \textit{“a good behavioral model has a Gini of 0.6 while a decent application model has a Gini of at least 0.3 because the data is scanty.”} 

Product and business analysts in banks expressed frustration with persistent gaps between Gini metrics and performance in real-world settings. A product lead who has worked for a large telco and now works in banking for over two decades shared his concerns.

\begin{quote}
    \textit{“So I have struggled in the past where you’d find a score card that always tend to have a high, let’s say a Gini of 80\% and 85\% [0.8 - 0.85]. But then when you get to performance at the end of the day, once you originate credit based on those scorecards, you find that you have defaults of  [between 30 to] 15\%. Is that expected ideally? Probably no.”} \\
    -- \textsf{[P4], Bank product lead}
\end{quote}

From the above, following the Gini formula,\footnote{Gini is derived from the AUC (Area Under the Curve of the Receiver Operating Characteristic) curve: Gini = 2 × AUC - 1; A Gini of 0 means the model has no discriminatory power (random), and Gini of 1 (or 100\%) means perfect separation.} the claimed 80-85\% should typically produce default rates of 3-6\% (compared to defaults of 10-30\%). The Gini for the reported default rates around 10-30\% was likely lower, ranging from negative values to at most 25\% (or 0.2). [P3] and [P4] suggested that data limitations have long plagued digital lending, speculating that credit data from CRBs and internal (bank) records were inadequate. While more data was spotlighted as the concrete solution to the gaps in credit scoring, P4 argued that even for large players like Safaricom, which have extensive data, their application scoring models also faced similar challenges as the rest of the industry. Ultimately, it was the “good” customers who were drawn on to boost the overall performance metrics of loan products. It was common knowledge that scale (users, loan books, data, etc.) would not only strengthen the predictive power of credit scoring but could help to rebalance loan portfolios despite known market realities (around default). Our participant, echoing what many others had said, developed this point more concretely:

\begin{quote}
    \textit{“The application scorecard in Kenyan market will yield a default of between 20 to 30\% for the mass market. If you lend, 20\% will not pay. The good thing is the 80\% who pay will always tend to pay afterwards [and] you’ll always form a repeat base, right? So your problem statement then will become do you have a scorecard that can minimize default from 20 to say 15\% or 10\% because you can’t bring it to lower than that right? [...] So you have to use the good customers to take care of the bad customers [...] [Players like Safaricom] have spent a lot of time to acquire good customers [whome they can] lend to without any scorecard. Well, they are good [customers], but then you still need to keep mining new customers, right? And that’s the value of the scorecard. But the question is at what cost?”} \\
    -- \textsf{[P4], Bank product lead}
\end{quote}

Often, the conversation went beyond Gini scores, returning to spaces of discretion, where business and lending decisions were made within the context of established social norms, relations, and obligations. Because cultures of managing and tracking credit scores were not (yet) part of everyday credit experiences, the work of interpreting and translating these scores relied on the social and cultural awareness of product and business managers (for example understanding of general loan default dynamics). According to P8, who had previously developed risk models abroad, the differences in credit cultures between Kenya and Western countries had to be put in context:

\begin{quote}
    \textit{“I think that [credit] scores literally mean nothing, because fundamentally, they do not really quantify risk accurately [...] If it’s my first time applying for credit, you know, it doesn’t mean that I’m a bad borrower because I don’t have any credit history. But also, you have to realize culturally here me not having [formal] credit is good.”}
\end{quote}

There were several examples of financial actors overlooking credit scores or considering them in part: P7 who works as an analyst at a Nairobi-based investment fund said that credit scores from CRB credit reports were \textit{“OK”} but did not necessarily reveal why an investment opportunity was not worth it. A lot more work was done to understand the entrepreneur at a more personal level. In another instance, P23 blamed these personal and relational practices for the \textit{“great contribution to the rate of the NPLs that we see [today].”} Borrowers (including those negatively listed at the CRB) draw on their relationships with directors and loan officers at SACCOs, especially church or community-based ones, to access loan facilities. The exceptions are the rapidly growing small-ticket loan books of digital lenders, where automated decision systems are applied at scale. 

Two takeaways emerge from this section. First, gaps between claimed model performance and observed outcomes were managed through justification, attribution of blame, and discretionary judgment that sometimes overrode model outputs. Second, when political or cultural imperatives collided with the market logics of digital lending, default rates surged, raising questions about whether credit scoring can function at scale without systematically excluding borrowers or relying on coercive enforcement mechanisms.


\section{Discussion}

The scenes and stories documented above are in dialog with prior HCI and human-centered ML contributions around the sociotechnical and institutional work of data science that makes ML systems function \cite{mao_how_2019,muller_how_2019,muller_forgetting_2022,passi_trust_2018}, the technical limits of predictive systems \cite{liu_bridging_2025,wang_against_2024}, the range of privacy and accountability considerations, and how they draw various interventions \cite{cooper_accountability_2022,coston_validity_2023,solove_privacy_2021}, and discretionary practices that are brought to bear when data and AI-driven systems encounter the world \cite{paakkonen_bureaucracy_2020,passi_data_2017,qadri_seeing_2022}. In our work, we observed the technical, institutional, and political practices of startups and other actors in the credit scoring worlds of Nairobi, where “risk” is the organizing technology and principle. Through ethnographic interviews and observation, we followed how expanded access to non-traditional or alternative forms of data prefigured how risk was constructed, legitimated, and evaluated in credit scoring. In our account, risk is specified in ever more granular terms, and in ways that at once disentangle the individual from their broader social relations while at the same time leveraging those relations to redefine and assess creditworthiness. In the first empirical section, we showed how the processes of collecting and classifying alternative data draw on legal and often questionable methods, yielding new types of data that nevertheless inherit dependencies from the contexts in which they are embedded. In the second section, we examined how risk was constructed using various technical and social definitions to support model design and development. These expanded risk definitions masked certain forms of uncertainty, leading to concerns about model performance and efficacy. In the third section, we highlighted the political and cultural contexts where human discretion appears, particularly when credit scoring models encounter complex, real-world situations. Collectively, our work demonstrates that risk and uncertainty are deeply entangled, and that making credit scoring work requires ongoing alignment as a two-way translation between models and the worlds they claim to measure. Below, we discuss how the performativity of alternative data \textit{names} and \textit{produces} differentiated effects of risk classifications; show how the entanglement of risk and uncertainty limits and challenges prediction; and discuss the important work of alignment, how it works, and for whom. We conclude with some questions for research and policy in high-stakes domains like credit scoring. 

\subsection{The performativity of alternative data}
As our findings show, alternative data prefigured outputs in algorithmic credit scoring. This is not simply because alternative data provide the inputs for defining and updating the risk categories that the models rely on, but also because these data are fundamentally \textit{performative}: they produce the effects that they name \cite{amoore_politics_2013}. One of the main claims encountered multiple times in our fieldwork was that expanded access to alternative data would enable startups and other fintech actors to mitigate fraud and risk, enhance automated decision-making, and widen a digital lender’s profit margins, while extending credit to groups otherwise excluded. Although these claims produced varying outcomes (as discussed in the next section), the new regimes of legibility they gave rise to, via devices, analytic techniques, and new forms of expertise, but also adjustments to borrower and consumer behavior enacted the worlds they were designed to measure. In this way, as \citet{bowker_biodiversity_2000} has emphasized, the world comes to be (partly) remade in the image of data: what is structured as countable, however skewed, creates a new world in which those counts are normalized and perhaps made normative (including through the assignment of values of ‘good’ and ‘bad’ behavior, ‘risky’ and ‘safe’). Accordingly, performativity is about who or what \textit{sees} and who or what \textit{is seen}, and the effects produced in this wake. The “who or what \textit{sees}” includes technical devices and institutional practices that facilitate the collection, analysis and operationalization of data, as well as the actors (practitioners or companies) who are granted this new mode of (in)sight \cite{moller_who_2020,sambasivan_everyone_2021,singh_seeing_2021}. The “what \textit{is seen}” are the specific types of data (and surrounding lifeworlds they represent) that can be classified and categorized, including as more or less relevant to prediction models \cite{jonas_friction_2019,muller_forgetting_2022,sambasivan_seeing_2021}. 

Throughout our findings, we show that the work of making alternative data is a social and technical accomplishment reliant upon: i) vast amounts of data accessed through legitimate and/or questionable means, and ii) discretionary judgments built around local knowledge of credit cultures and behaviors. Because of Safaricom’s monopoly control over rich real-time data infrastructure, external access was significantly limited. In this context, startups made technical and moral claims to justify their approaches to collect potentially unlimited amounts of trace data generated by Safaricom (and other actors, including banks and commercial stores) which was believed to capture a rich and mineable symbolic surrogate for the vital detail of an individual’s or an organization’s character and life \cite{zuboff_age_1988}. The startups in our study built this alternative data through technical workarounds and reverse-engineering – for example, the practice of “SMS scraping” described in the findings. These workarounds were made possible through the use of SDKs and APIs \cite{reardon_50_2019}, which enabled the startups to not only automatically collect data from their own apps (i.e., loan apps they developed) but also from a wider range of banking, CRB and fintech apps to which they had “third-party” access \cite{mutungu_privacy_2021}. Sources such as “public-private” partnerships and informal data-sharing agreements enabled access to wider and more diverse data ranging from agriculture to sports betting. While these data collection processes have become normalized as standard practice, they operate at or beyond the margins of legality, complicating what is considered ethical or normative, and inclusive or predatory. Despite the use of “right permissioning” systems to support data collection (several of our participants were quick to invoke the necessity of complying with Kenya’s Data Protection Act), the extent of privacy intrusion and surveillance, including through backgrounded processes, meant that data was collected and used well beyond the purposes it was originally intended for (and certainly beyond the bounds of any reasonable application of the contextual integrity standard \cite{nissenbaum_contextual_2019,lauer_creditworthy_2017,thylstrup_politics_2022}).

Where the modes of classification and analysis drew on distinctions between various forms of data – static and dynamic – they widened possibilities of misinterpretation and the automation or reproduction of bias in credit scoring. Where static data specifying individual demographic information (age, gender, etc.) was mixed with dynamic data capturing large volumes of user data including real-time customer transactions, user device IDs, and connections between users and their broader social networks \cite{amoore_politics_2013,fourcade_ordinal_2024}, the ability to infer and discriminate on the basis of what in the US context is described as “protected attributes” \cite{barocas_big_2016} radically expands (in the Kenyan context, data associated with these attributes are easily accessible through app-based KYC processes and even the government’s fee-based Integrated Population Registration Service (IPRS) API \cite{breckenridge_failure_2019}). Since alternative data facilitated access to dynamic and temporal aspects of everyday user behavior and transactions, it was seen by our interlocutors as a source of both positive (e.g., active consumption) and negative (e.g., gambling) risk signals. However, the startups studied were so intense in their search of risk signals that benign and ordinary behaviors and social practices were classified as “risky.” For example, Moonshot’s data science lead said that they assigned unique device IDs to a borrower’s phone to analyze whether other people were using the same phone to borrow. By individualizing phone usage, Moonshot’s risk score separated a user from socially and culturally significant practices of phone sharing, which are common in rural areas and among low-income earners in urban areas, whom fintech actors purported to financially include. These practices have been widely documented in HCI and ICTD literature \cite{bidwell_decolonising_2016,burrell_evaluating_2010,donner_after_2015,wyche_dead_2012,wyche_understanding_2019}. When these operations are connected to meaningful outcomes in the lifeworld, as the practitioners’ concerns around ``gaming'' suggest, users begin to reshape their actions and practices. Slowly and subtly, \textit{what is counted} becomes \textit{what counts}.

\subsection{Risk and the (sociotechnical) limits of prediction} 
In our findings, risk is not only complex and dynamic but also inextricably connected to uncertainty (as we will demonstrate in the next paragraphs), complicating whether “risk” can be substantively bounded and rendered computable in predictive systems. Our interlocutors in data science teams acknowledged that despite their endless pursuit of more and more alternative data, ML uses narrow and historical snapshots of (risk) data to make predictions. This is the fundamental design (and limitation) of ML systems that several studies of situated data science practice have extensively documented, demonstrating how data science practitioners recognize and manage these limitations \cite{malik_hierarchy_2020,muller_how_2019,passi_trust_2018}. Through continuous improvements in data quality and coverage and model retraining, ML systems can be updated to be “good enough” \cite{bialski_middle_2024}. Across our sites, alternative data served as proxies of risk – based upon borrowers’ personal and social lives – especially in the absence of reliable borrower credit data (the so-called “thin-file” or “credit invisible” borrowers). Additionally, the dynamic and temporal quality of alternative data enabled data science teams to update their risk assumptions and subsequently credit scoring models. As we have observed in our findings, features associated with psychological and sociological aspects of the users’ contexts were extracted from static and dynamic datasets. For example, features were engineered to capture whether a customer saved money or indulged in sports betting. While hundreds (or even thousands) of features could be generated, the goal was ultimately to \textit{reduce} the number of features to a handful, between 10 and 20. \textit{Risk reduction} techniques such as correlation analysis were widely used to question human intuitions that could generate spurious results, explaining the significance of particular features. Although some data science teams assumed that practices such as saving money were positively correlated with creditworthiness, and sports betting negatively correlated, several participants reflected about moments when these assumptions did not hold to be true (and most of the insights they landed on in these “non-scalable manual checks” became part of proprietary and secret techniques that they said gave them advantage over their competitors).

Despite the workarounds (and everyday routines) adopted in data science practice (correlation analysis, explainability tools, data updates, model retraining, etc.), the relationship between risk and prediction in ML systems is often tenuous, especially in high-stakes decision domains. More recently, research in HCI and adjacent fields has found that predictive systems often amplify the very risks they aim to mitigate \cite{saxena_rethinking_2023,wang_against_2024}. Some of this work questions the epistemological foundations of risk, highlighting the various factors – temporal, systemic, environmental, community-level, etc. – that can influence and distort risk predictive outcomes \cite{saxena_rethinking_2023}. Our findings show that risk both masks and is entangled with at least three forms of uncertainty. 

First, \textit{epistemic} uncertainty is associated with what can be known and hence measured \cite{knight_risk_1921}. In many ways, the pursuit of alternative data was an attempt not only to grapple with this uncertainty but also to expand the horizons of what could be known. However, despite the seeming totality of alternative data, infrastructural dependency gaps, some of which were known, inhibited these efforts \cite{jackson_understanding_2007}. For large players like Safaricom, repositories of data were so fragmented that it was near impossible for them to leverage these data in predictive models: an interlocutor intimated that costs were prohibitively high to substantiate the transformation and usage of such data. Relatedly, the overreliance on identity infrastructure, such as SIM cards, for sourcing and verifying borrower identities exposed several digital lenders to the limitations of these data infrastructures. While SIM cards have democratized access, their fungibility means that data profiles frequently get mixed up, undermining eventual credit scoring: when a SIM card that was previously owned by a loan defaulter is circulated in the market, the new SIM holder can accidentally take on the burden of the loan (including instances of blacklisting at the CRB). 

Second, \textit{modeling} uncertainty entails assumptions about the design of ML models and claims about how these models work in the real world \cite{kou_dead_2025}. This uncertainty was most prominent in aspects where our interlocutors made modeling decisions based on what they had observed in the data and how local markets operated. One example is the use of credit limits instead of credit scores (as in the US and most parts of the world). Rather than rejecting dubious applicants, borrower-specific credit limits were assigned from the onset (credit floors were as low as US\$0.8), only adjusted with more data on borrower behavior. However, the processes of reducing risk to manageable levels were confronted by dynamic and temporal shifts within datasets and scoring models, and of course, everyday borrower behavior. 

Third, \textit{contextual} uncertainty concerns the discretionary power exercised over model predictions \cite{paakkonen_bureaucracy_2020}. In the face of rising NPL rates and poorly calibrated Gini coefficients, the interlocutors who worked closely with business analysts and product teams said that discretion was exercised to justify the supplementation of predictive scores with some other kind of business rule or political reasoning. Practitioners who used these systems in credit decisions drew on their experience and expertise to make judgment calls, especially when credit models failed to meet stated claims. Although feedback loops between model developers and credit practitioners explored possibilities for improvement (more data, hybrid scores, etc.), there was exasperation about whether uncertainty could be fully addressed in light of overlapping workflows and differences in risk appetites in the digital lending space.

Altogether, data science and ML practitioners recognize the design limitations of machine learning and employ a range of techniques and tools to mitigate them. However, the relationship between risk and uncertainty makes it difficult to substantively claim predictive accuracy, validity, and consistency – especially where incomplete data, varied modeling assumptions, and discretionary practitioner interventions are difficult to reconcile.

\subsection{Alignment for what, whom, and when?} 
We have built on \citet{fujimura_constructing_1987} and HCI scholars \cite{dourish_allure_2021,fahimi_articulation_2024} to define alignment as the technical and discursive work that supports coordination and collaboration across multiple levels of practice, united by commonly shared problems and interests. Prior work in HCI has primarily examined one direction of this alignment: making models acceptable to organizational and social worlds through explainability, justification, and negotiation \cite{arya_one_2019, fair_isaac_corporation_machine_2018, passi_trust_2018}. AI safety discourse \cite{ahmed_field-building_2024,ji_ai_2025} (which purports to bring human values and guardrails into large language models – in effect, making \textbf{“models safe for the world”}) shares this one-directional framing but treats it as a normative constraint problem rather than an analytic account of coordination. In contrast, alignment as used here emphasizes the two-way translation between data/models and worlds, which includes at least some dimension, as we have already seen, of making \textbf{“worlds safe for models.”} In our findings, credit scoring – and the technologies, institutions, and worldviews underlying it – was made amenable (safe, inclusive, responsible, etc.) to the world; while the world – as borrowers were disciplined into ``good'' debtors and ordinary practices like phone sharing were pathologized as risky – was made amenable to credit scoring.

A part of this work was achieved through the efforts, practices, and perceptions of the varied actors in our study, along with stories around the importance of democratizing access to credit through credit scoring technologies. The Moonshot Extenscore® launch was celebrated as a triumph of inclusion and responsibility, a social act by a technical team of data scientists and experienced analysts who mastered risk by creating a hybrid score that trounced competitors in the market. This supposedly illustrated the power of alternative data to extend credit to the underserved and marginalized of society. The institution of privacy regulations (the Data Protection Act and attendant regulations) and appointment of an officer to overlook data violations in digital lending (i.e., Office of the Data Protection Commissioner) were perceived as efforts that promoted the efficacy and safety of algorithmic credit scoring and its orderly reception in the market. Participatory processes in the private and public sectors, from user surveys and focus groups to the inclusion of diverse voices in product research and development to invitations from the Central Bank for public comments on draft proposals and pending regulations sought to bring a modicum of regulatory oversight and assurance to digital lending products and practices.

At the same time, we saw evidence of worlds being remade in the ``image'' of models – adjustments to behavior, evaluation metrics, practice and opportunity that stemmed from the widening presence and significance of credit scoring. Due to high NPLs, borrowers in many instances were castigated and blamed for being serial defaulters (at least according to the PS in the Hustler Fund scene). Fintechs and banks sought new mechanisms to include but also exclude potential borrowers, effectively screening in and screening out new (and old) customers for their services.  Much of this work was disciplinary in nature, enforcing propriety and compliance by the threat of financial consequence. Borrowers were called on to be “good” debtors: “your behavior is your property.” While borrower-specific data was meant to be shared with CRBs to harmonize credit information sharing among regulated financial institutions such as banks, CRBs were often instead used as sites of blacklisting, punishing supposed cases of non-repayment (including for infractions under US\$2) through the loss of access to future financial resources, carrying in turn key implications for sectors from housing to education to healthcare. Another aspect of this work also entailed rebalancing of performance in loan portfolios in the face of rising NPLs, as well as rationalizing away failure. Digital lenders, for example, drew on “good” customers to offset the effects of “bad” borrowers (who were charged high interest rates anyway), thus sustaining the growth of their loan books. Relatedly, despite registering the highest default rate in the market at 68.3\%, the Hustler Fund’s promoters defended its poor performance, arguing it is a policy intervention rather than a profit-driven initiative.

\section{Conclusion}

In this paper, we have examined the sociotechnical and institutional work of machine learning and data science in developing credit scoring systems under local constraints around data access, enduring credit cultures, and the shifting policy and regulatory contexts of Kenya. We argued that while startups, banks, telcos, CRBs, and other fintech actors redefine risk through expanded access to alternative data, risk is fundamentally entangled with uncertainty in epistemic, modeling, and contextual forms, complicating how models are constructed, deployed, and evaluated in real-world contexts. We extended work on alignment in HCI (and offered a counterpoint to AI safety discourse) to argue that credit scoring is accomplished through a two-way translation where models are made safe for worlds, while those worlds are reshaped to fit model assumptions. This alignment work brings temporary stability and a kind of provisional order, but with uneven consequences for privacy, surveillance, and the terms of financial inclusion. At the interface of this two-way translation are contradictions which demand normative, practical, and sometimes technical responses. Who is responsible for reconciling legal gaps where the law protects fundamental privacy rights while credit information frameworks are silent on the legality (and misuses) of alternative data? When and how should normative questions around privacy be balanced with developmentalist aspirations to access and inclusion? In the absence of punitive measures (such as negative listing at CRBs), can these models function effectively at scale in complex, real-world situations? And perhaps above all: what is risky, what is uncertain, and who should bear the costs and weight of each?

\begin{acks}
We would like to thank Miriam Fahimi for thoughtful comments on an earlier draft, Amiel Bize and Kevin Donovan for their sustained engagement throughout the project, and the reviewers for their patient and generous feedback. We also thank our interlocutors who shared their time, experiences, and expertise with us.
\end{acks}

\bibliographystyle{ACM-Reference-Format}
\bibliography{references}

\end{document}